\documentclass[runningheads]{llncs}

\usepackage{graphicx}
\usepackage{hyperref}

\usepackage{multirow}
\usepackage{tabu}
\usepackage{bbm}
\usepackage{diagbox}
\usepackage[english]{babel}
\usepackage{comment}
\usepackage{array}
\usepackage{amstext}
\usepackage{makecell}
\usepackage{caption}
\usepackage{tablefootnote}
\usepackage{babel}
\usepackage{booktabs} 
\usepackage{color}
\usepackage{amsmath}
\usepackage{wrapfig}
\usepackage{amssymb}
\usepackage[ruled,vlined]{algorithm2e}
\usepackage{subfigure} %需要使用的宏包

\SetKwInput{KwInput}{Input}               
\SetKwInput{KwOutput}{Output}         
\SetKwInput{KwParameter}{Parameter}    
\SetKwInput{KwInitialize}{Initialize}          

\newif\ifdraft
\draftfalse
\drafttrue

\ifdraft
 \newcommand{\PF}[1]{{\color{red}{\bf PF: #1}}}
 
 \newcommand{\MS}[1]{{\color{green}{\bf MS: #1}}}
 
 \newcommand{\ZD}[1]{{\color{violet}{\bf ZD: #1}}}
 
 \newcommand{\YH}[1]{{\color{blue}{\bf YH: #1}}}
 
 \newcommand{\SPE}[1]{{\color{orange}{\bf SS: #1}}}
 
 \newcommand{\WJ}[1]{{\color{orange}{\bf WJ: #1}}}
 
  \newcommand{\JY}[1]{{\color{blue}{\bf JY: #1}}}
 
  \newcommand{\placeholder}[1]{{\color{green}{placeholder: #1}}}
\else
 \newcommand{\PF}[1]{}
 
 \newcommand{\KY}[1]{}
 
 \newcommand{\MS}[1]{}
 
 \newcommand{\ZD}[1]{}
 
 \newcommand{\YH}[1]{}
 
 \newcommand{\SPE}[1]{}
 
 \newcommand{\WJ}[1]{}
 
 \newcommand{\JY}[1]{}
 
 \newcommand{\placeholder}[1]{}
\fi

\begin{document}

\title{Topology Repairing of Disconnected Pulmonary Airways and Vessels: Baselines and a Dataset}

\titlerunning{Topology Repairing of Pulmonary Airways and Vessels}

% \author{Ziqiao Weng, Jiancheng Yang, Dongnan Liu, Weidong Cai}
% \institute{School of Computer Science, University of Sydney, Sydney, Australia}

\author{Ziqiao Weng\inst{1} \and
Jiancheng Yang\inst{2}\thanks{Corresponding author: Jiancheng Yang (jiancheng.yang@epfl.ch).} \and
Dongnan Liu\inst{1} \and
Weidong Cai\inst{1}}
\authorrunning{Z. Weng et al.}
\institute{School of Computer Science, University of Sydney, Sydney, Australia \and
Computer Vision Laboratory, Swiss Federal Institute of Technology Lausanne (EPFL), Lausanne, Switzerland
}
% %
\maketitle              % typeset the header of the contribution
%

%%%%%%%%% ABSTRACT
% !TEX root = ../top.tex
% !TEX spellcheck = en-US

\begin{abstract}

Accurate segmentation of pulmonary airways and vessels is crucial for the diagnosis and treatment of pulmonary diseases. However, current deep learning approaches suffer from disconnectivity issues that hinder their clinical usefulness. To address this challenge, we propose a post-processing approach that leverages a data-driven method to repair the topology of disconnected pulmonary tubular structures. Our approach formulates the problem as a keypoint detection task, where a neural network is trained to predict keypoints that can bridge disconnected components. We use a training data synthesis pipeline that generates disconnected data from complete pulmonary structures. Moreover, the new Pulmonary Tree Repairing (PTR) dataset is publicly available, which comprises 800 complete 3D models of pulmonary airways, arteries, and veins, as well as the synthetic disconnected data. Our code and data are available at \url{https://github.com/M3DV/pulmonary-tree-repairing}.

\keywords{pulmonary airways \and pulmonary vessels \and tree structure repairing \and geometric deep learning \and shape analysis.}

\end{abstract}

%%%%%%%%% BODY TEXT
% !TEX root = ../top.tex
% !TEX spellcheck = en-US

\section{Introduction}

Pulmonary diseases pose significant health risks, and computed tomography (CT) analysis of pulmonary airways and vessels has become a valuable clinical tool for revealing tomographic patterns \cite{qin2021learning,zhang2023multi}. Precise representation of the airway tree is essential for quantifying morphological changes, diagnosing respiratory disorders such as bronchial stenosis, acute respiratory distress syndrome, idiopathic pulmonary fibrosis, chronic obstructive pulmonary disease (COPD), obliterative bronchiolitis, and pulmonary contusion, as well as for virtual bronchoscopy and endobronchial navigation in surgery \cite{qin2021learning,fetita2004pulmonary}. Furthermore, accurate modeling pulmonary arteries and veins improves computer-aided diagnosis of pulmonary embolism, chronic pulmonary hypertension \cite{rahaghi2016pulmonary,Wittenberg}, and lobectomy/segmentectomy~\cite{saji2022segmentectomy,zhao20183d,zhao2023invasiveness}.

% !TEX root = ../top.tex
% !TEX spellcheck = en-US

\begin{figure}
    \centering
	\includegraphics[width=0.85\linewidth]{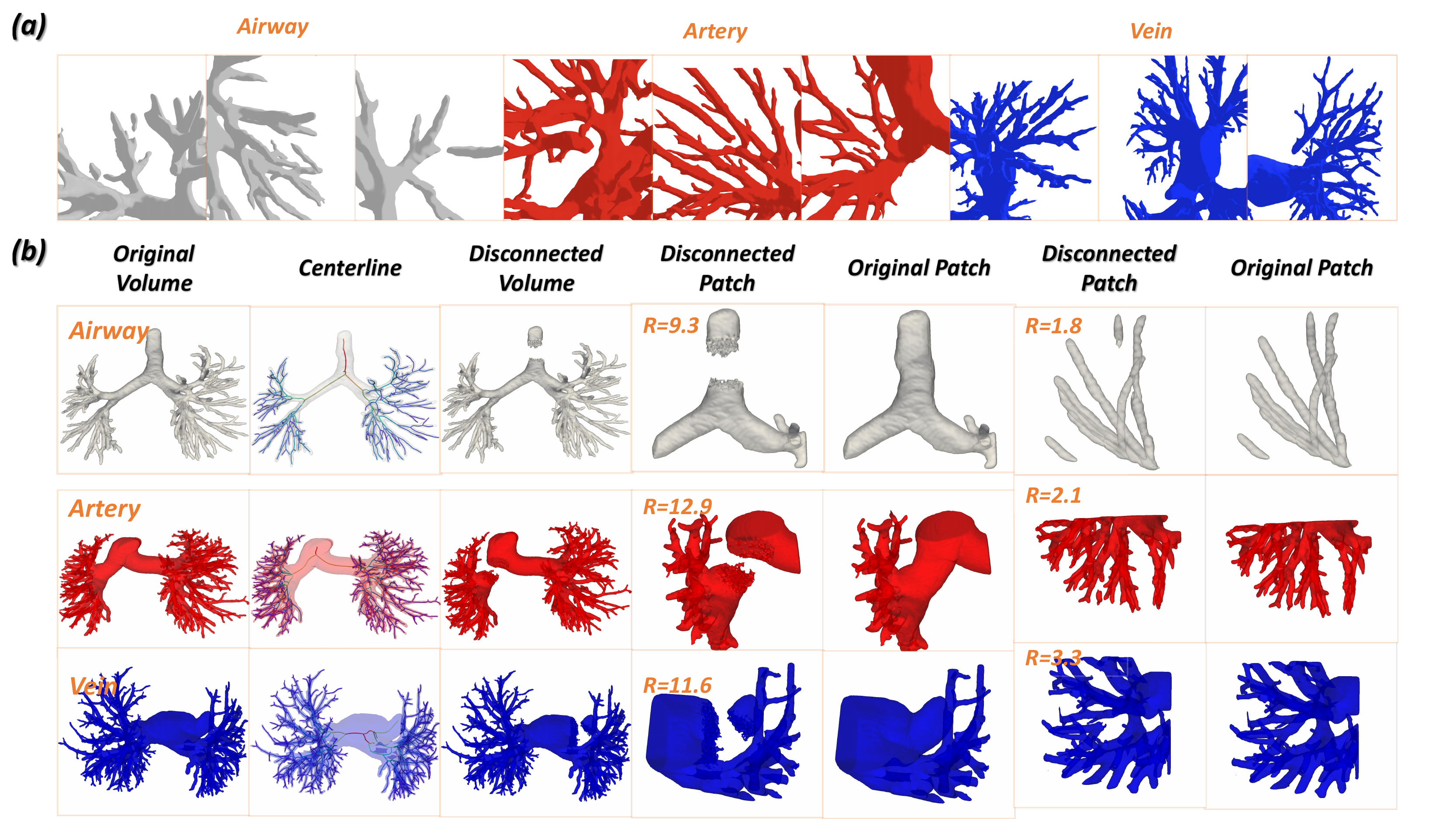}
	\caption{\textbf{Visualizations of the PTR dataset.} \textbf{(a)}: Examples of disconnected predictions from nnU-Net. \textbf{(b)}: From left to right: original volumes, centerlines extracted by Vesselvio (with different edge radii indicated by colors), disconnection synthesis, subvolume views of disconnected parts, and corresponding original parts. The volumes have been smoothed for better visualization. The average edge radius, measured in micrometers, is denoted by "R" in the figures. Airways, arteries, and veins are respectively colored in gray, red, and blue.}
	\label{fig:data_generation}
	\vspace{-20px}
\end{figure}

In recent years, deep learning methods have spawned research on airway and vessel segmentation. Convolutional neural networks (CNNs) have been widely employed in various existing studies to learn robust and discriminative features for automatic airway/artery/vein segmentation \cite{qin2021learning,andongW,tetteh2020deepvesselnet,DeepVessel,garcia2021automatic,park2021deep,zhang2023multi}. However, accurately reconstructing complete airway or vessel tree branches remains a major challenge. Current state-of-the-art segmentation models, such as nnU-Net \cite{isensee2021nnu}, still suffer from inadequate precision due to the minute scale and scattered spatial distribution of peripheral bronchi and vessels, which causes a severe class imbalance between the foreground and background, leading to degraded segmentation accuracy. The implications of such degraded performance can have negative consequences on clinical judgments and diagnoses, as it can lead to disconnections of pulmonary tubular structures of airways or vessels, as depicted in Fig.~\ref{fig:data_generation}, potentially impeding accurate medical assessments.

In this paper, we formulate the problem of disconnected pulmonary tubular structures as a key point detection task. The primary objective is to repair the topology structures of two disconnected components by accurately identifying the centers of the disconnected parts located at both ends of the components. Endpoints corresponding to the broken centerline of the pulmonary tubular structure are treated as two key points. The identification of these key points is critical in recognizing disconnections in pulmonary tubular structures for diagnosing pulmonary diseases, which has significant research implications. To address this issue, we propose a training data synthesis pipeline that generates disconnected data from complete pulmonary structures. We further explore the training strategy and thus build a strong basline based on 3D-UNet to predict the key points that can bridge disconnected components.
Our contributions can be briefly summarized as follows:

\begin{itemize}
  \item \textbf{A novel formulation of a practical research problem}: We have formulated the problem of pseudo disconnection pulmonary tubular structures as a key point detection task, which is a significant contribution to the field as it has not been extensively explored before;
  
  \item \textbf{An effective yet simple baseline with efficient 3D-UNet}: We propose a two-channel 3D neural network that efficiently identifies key points and bridges disconnected components. Our model demonstrates decent performance, providing a strong baseline for future studies.
  
  \item \textbf{An open-source benchmark}: To evaluate the proposed model, we have constructed a new pulmonary dataset named Pulmonary Tree Repairing (PTR), and designed proper metrics for performance examination. This dataset will be publicly available soon and will enable reproducibility and comparison of future studies in this field.
\end{itemize}

% !TEX root = ../top.tex
% !TEX spellcheck = en-US

\section{Method}

In this section, we present a comprehensive analysis of our approach for detecting pulmonary tubular interruptions as a keypoint detection task. We start by formulating the problem, followed by a description of the data simulation process used to construct the dataset. The dataset construction process is explained in detail to provide insight into the methods used for generating realistic data samples. We then introduce the simple two-channel 3D-UNet, and describe its architecture, key features, training objective, and implementation details.  

\subsection{Problem Formulation}

Segmentation of thoracic tubular structures, such as airways and vessels, from lung Computed Tomography (CT) scans is vital for diagnosing pulmonary diseases. Over the years, various deep learning-based segmentation methods have demonstrated the potential of Convolutional Neural Networks (CNNs) in handling this task. However, accurately segmenting pulmonary airways, arteries, and veins without interruption remains challenging due to the unique properties of the thoracic tubular structure. The trachea and blood vessels constitute only a small fraction of the whole thoracic CT image, which leads to severe class imbalance between the tubular foreground and background, hindering 3-D CNNs learning from sufficient supervisory signals \cite{andongW}. Moreover, airways and vessels are complex tree-like structures with numerous bifurcations and branches of various sizes and lengths, making it difficult for CNNs to capture fine-grained patterns without encountering memory/parameter explosion and overfitting \cite{qin2021learning}. Segmentation networks often produce unsatisfactory predictions, resulting in disconnection or interruption of the estimated tubular structure, which could affect clinicians' judgment in clinical practice. Therefore, identifying the location of disconnections is of great research importance. In this paper, we have formulated the problem as a key point detection task, with the two endpoints of the interrupted centerline of the tubular structure serving as the two key points. We aim to use neural networks to predict the location of the disconnection part of vessels/airways, which has significant research implications.

Keypoint detection is a popular computer vision technique for identifying object parts in images, with applications ranging from face recognition, pose estimation to medical landmark detection~\cite{sun2019deep,xiao2018simple,WangJD,yao2021one}. Heatmap regression has emerged as a standard approach for keypoint detection, where ground-truth heatmaps are generated for each keypoint using a Gaussian kernel \cite{yu2021heatmap,LuoSWAHR}. The network outputs multi-channel heatmaps, with each channel corresponding to a specific keypoint. Our work adopts this approach for detecting two keypoints located at the endpoints of interrupted airway/vessel. 

\subsection{Training Data Synthesis}

We generated synthetic data from lung CT scans with carefully annotated pulmonary airways, arteries, and veins, as no public medical dataset was available for the task at hand. The synthetic data simulates the scenario of vascular/trachea disconnection and serves as a benchmark dataset for the keypoint detection task. To generate the data, binary masks of the tubular structures were extracted from 800 CT scans \cite{Kuang2022WhatMF}, and VesselVio software \cite{bumgarner2022vesselvio} was used to identify the centerlines of binarized airway/vessel volumes and create tree-like graphs. Random sampling was performed to select a branch of the vessel or airway, and two keypoints were sampled along the pre-extracted centerline. The keypoints were then subjected to morphological operations (from SimpleITK Python library \cite{SimpleITK1,yaniv2018simpleitk}) to create near-true vascular disconnections. The resulting keypoints were labeled $\mathit{KP_{1}}$ and $\mathit{KP_{2}}$, and the data was visualized in Fig.~\ref{fig:data_generation}. It is important to note that discontinuities in real-world scenarios are mainly observed in thinner blood vessels. Due to the random sampling process and the prevalence of small branches within the entire tubular structure, the generated discontinuities are predominantly manifested in small blood vessels. Including the subfigures in Fig.~\ref{fig:data_generation} (b) aims to clearly illustrate the visual appearance of these generated discontinuities.

\subsection{The Keypoint Detection Network}

\begin{figure}[tb]
    \centering
	\includegraphics[width=0.8\linewidth]{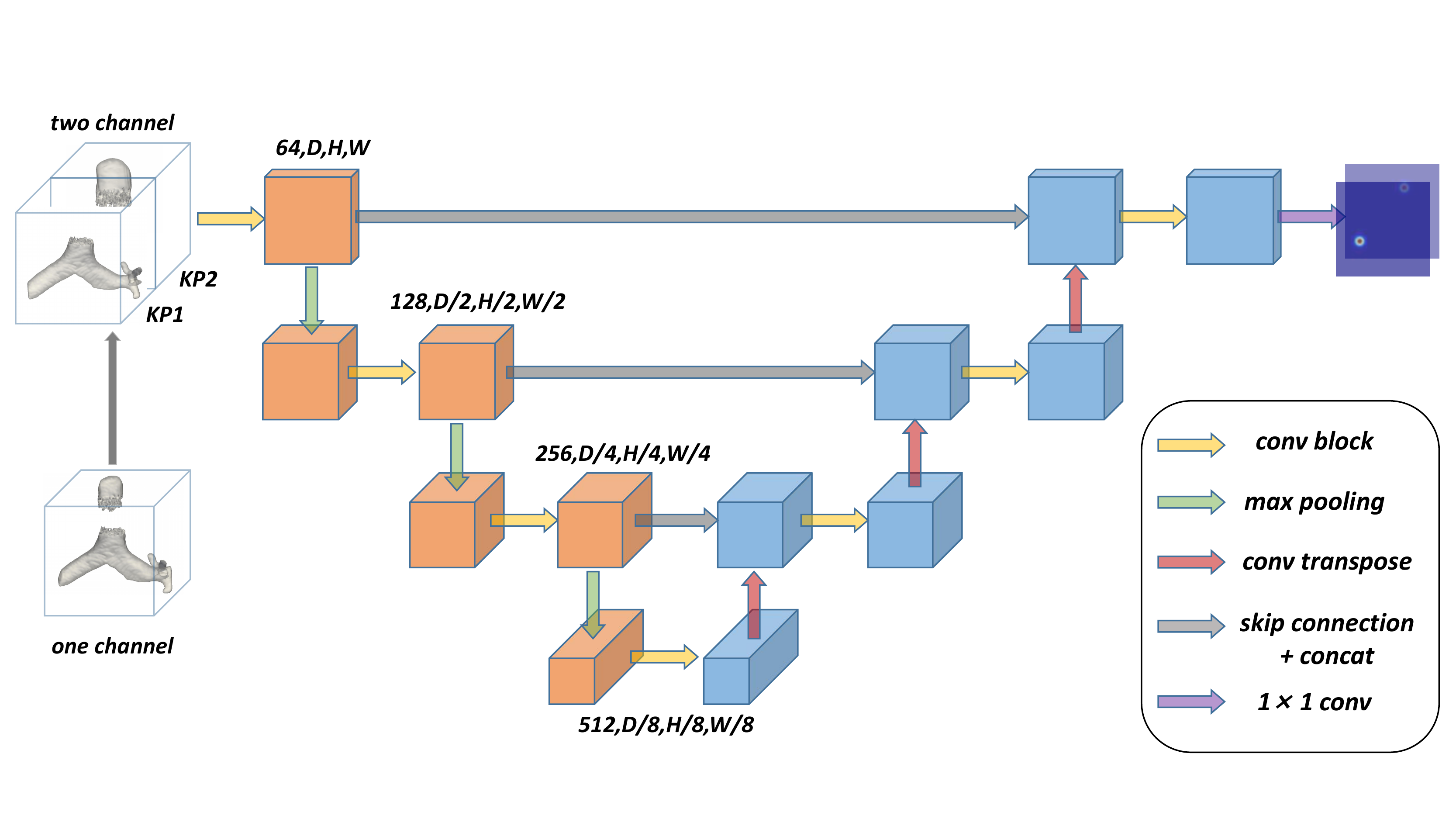}
	\caption{\textbf{Keypoint Detection Network.} Given two disconnected components, the 3D-UNet outputs two heatmaps corresponding to $\mathit{KP_{1}}$ and $\mathit{KP_{2}}$.}
	\label{fig:network}
	\vspace{-15px}
\end{figure}

The framework and training pipeline of our network are depicted in Fig.~\ref{fig:network}. In the following sections, we will introduce each component in detail.

% !TEX root = ../top.tex
% !TEX spellcheck = en-US

\paragraph{Data Sampling.} 

The generated raw data is too large for training the network due to the high-resolution nature of CT scans, which have dimensions of $512\times 512$ for the x-y plane and variable dimensions for the z plane. Directly feeding the entire 3-D volume into the network can cause significant memory overhead and slow down the training convergence, especially with limited computing resources. Therefore, we crop a subvolume with a size of $80\times 80\times 80$ around where the disconnection occurs from the original volume. Specifically, since the location of interrupted blood vessels cannot be known in advance and the small connected component where $\mathit{KP_{2}}$ is located can be found using morphological operations, we randomly select a point in that small object as the center point of our subvolume. This approach also serves as a new form of data augmentation. For each selected branch in an original volume, we randomly crop one subvolume for training purposes and three subvolumes for validation and testing.

\paragraph{Network Design.} 
We propose an encoder-decoder network that is based on the widely used 3D U-Net architecture. As depicted in Fig.~\ref{fig:network}, the inputs to the network are obtained by cropping subvolumes of the same size as the original volume. The first input contains only $\mathit{KP_{1}}$ and its connected component in the whole volume but is presented in a subvolume view. The second input exclusively comprises the small vessel/airway segment of $\mathit{KP_{2}}$. The output heatmaps of the two keypoints correspond to the $\mathit{KP_{1}}$ input and $\mathit{KP_{2}}$ input, respectively, which avoids learning ambiguity. The 3D U-Net is a neural network architecture that features three encoder and three decoder stages. Each stage includes a convolution block and a downsampling or upsampling layer. The convolution block consists of two convolution layers, each using a kernel size of ${3\times 3\times 3}$, followed by batch normalization and rectified linear unit (ReLU) activation.

The network receives two binarized subvolumes $\boldsymbol{I} \in \mathbb{R}^{2 \times D \times H \times W}$ as inputs, where $D, H, W$ represent the spatial dimensions of the cropped volume. In this study, we decided to set the output heatmaps to the same size as the inputs, without downsampling, in order to avoid the loss of coordinate accuracy. 

In the neural network design phase, we prioritized formulating the problem, constructing an open-source dataset, and proposing a comprehensive training and testing pipeline. We refrained from incorporating sophisticated modules, such as attention mechanisms, transformer blocks, or distillation, and fine-tuning hyper-parameters. Hence, we do not delve into detailed network architecture design in this paper. However, we obtained promising results using a simple two-channel 3D-UNet model and explored various training techniques. Our work lays a solid foundation for future researchers to improve upon our findings by incorporating advanced techniques and innovative modules.

\paragraph{Loss Function.} 
We adopt the state-of-the-art keypoint detection framework to represent the problem as heatmap estimation, where the coordinate with the highest confidence in each heatmap of $\boldsymbol{H} \in \mathbb{R}^{k \times D \times H \times W}$ corresponds to the location of the $k$th keypoint. The ground-truth heatmaps are generated by placing a 3D Gaussian kernel at the center of each ground-truth keypoint location. For simplicity, we define the Keypoint Mean-Squared Error (KMSE) loss function as follows:
\begin{equation}
    \mathcal{L}_{kmse}=\frac{1}{K} \sum_{\mathit{k}=1}^{K} \delta \left ( {V_k}> 0 \right ) \cdot \left \| H_k-\hat{H_k} \right \|_2^{2},
\end{equation}
where ${H_k}$ and $\hat{H_k}$ refer to the ground-truth and the predicted heatmaps for the $k$th keypoint, and $K$ is fixed to 2 in our study. To reduce memory cost, we limit the size of subvolumes to ${80 \times 80 \times 80}$, which may result in invisible keypoints if the branch is long and the two keypoints are too far apart. Here, ${V_k}$ indicates the visibility of the ground truth keypoint, where ${V_k}=1$ and $\delta({V_k})=1$ if the keypoint is visible, and vice versa. 

\paragraph{Implementation Details.} 
During the training phase, we employ a sampling strategy that randomly crops one volume, which introduces data augmentation, mitigates overfitting, and ensures training convergence. To reduce testing time and enable fair comparisons between models, we generate and save three random crops for each vessel branch during validation and testing. The size of the ground-truth heatmaps is ${80\times 80\times 80}$, and the sigma of the 3D Gaussian kernel used to generate them is set to 2.5. All networks were trained using AdamW optimizer with a learning rate of 0.0001 and beta hyperparameters of 0.5 and 0.999. The training was performed on a single NVIDIA 3090ti GPU with a batch size of 16. PyTorch framework was used for implementation, and early stopping strategy was adopted to prevent overfitting. To speed up training, we initialize the artery and vessel models with the trained airway model. We combined artery and vein training data to increase the training samples and reduce the training time. 

\subsection{Model Inference}

Models trained using our proposed training paradigm may not be directly applicable to real-world data due to several assumptions made during training. Specifically, during training, we assume that the interrupted segmentation mask consists of only two continuous components representing $\mathit{KP_{1}}$ and $\mathit{KP_{2}}$, and that the location of $\mathit{KP_{2}}$ is known a priori, which is used to randomly crop subvolumes. Additionally, we limit the subvolume's size to ensure efficient training. However, in real-world scenarios, the location of $\mathit{KP_{1}}$ and $\mathit{KP_{2}}$ components is unknown, and there may be small disconnected objects and noises scattered throughout the volume's entire original size. The only prior knowledge available is that $\mathit{KP_{1}}$ is located in the volume's largest connected component (i.e., the main vessel/airway), and $\mathit{KP_{2}}$ is in one of the small isolated components. To address this issue, we have developed an algorithm that bridges the gap between model training and inference, and accurately predicts disconnections in real-world situations. The pseudo-code of the inference algorithm is detailed in the supplementary materials.

% !TEX root = ../top.tex
% !TEX spellcheck = en-US

\section{Experiments}

\subsection{Datasets}
A total of 800 CT scans with annotations of pulmonary airways, arteries, and veins are utilized to construct our dataset. The CT scans from multiple medical centers are manually annotated by a junior radiologist and confirmed by a senior radiologist~\cite{Kuang2022WhatMF}. The data is divided into training, validation, and test subsets with a ratio of 7:1:2. Each CT scan is pre-processed into three binarized volumes of airways, arteries, and veins. Subsequently, 30 distinct branches per volume were randomly selected for each binarized volume under specific criteria to create 30 volumes with vascular interruptions. {The Pulmonary Tree Repairing (PTR) dataset includes 3D models represented by binarized ground-truth segmentation masks, centerlines, disconnected volumes, and a corresponding json file for each subject. The json file contains comprehensive information, such as the coordinates of bifurcations, endpoints, and all points along each branch, capturing diverse characteristics specific to each blood vessel.} Note that the keypoint detection of airways, arteries, and veins disconnection is treated as three independent tasks, with each task having a dataset size of $800\times 30$. The results are optimized on the validation set and reported on the test set.

\subsection{Evaluation Metrics}
Based on the definition of Object Keypoint Similarity (OKS) in pose estimation tasks, we have adapted this metric to align with the features of our dataset. Our modifications to the OKS are reflected in the following metric formulations:

\begin{equation}
    \mathit{OKS}_{k}= \exp \left \{ -d_{k}^{2}/2S\lambda ^{2} \right \},
    \mathit{E}_{d_{k}}= \exp \left \{ -d_{k}^{2} \right \},
\end{equation}
Here $d_{k}$ is the Euclidean distance between the predicted keypoint and the ground-truth keypoint, along with the vessel volume $S$ of the corresponding branch. To maintain a consistent scale for OKS, we have introduced $\lambda$, a constant which we set to 0.2. $\mathit{OKS}_{k}$ refers to the $\mathit{OKS}$ of $k$th keypoint ($k=2$ in our study).
\begin{equation}
    \mathit{OKS}_{i}=\frac{\sum_{k}{\mathit{OKS}_{k}\cdot\delta \left({V_k}>0\right)}}{\sum_{k}{\delta \left({V_k}>0\right)}},
\end{equation}
where $\mathit{OKS}_{i}$ denotes the $\mathit{OKS}$ of $i$th sample and ${V_k}$ is the visibility flag. 
\begin{equation}
    \mathit{AP}^{\tau}=\frac{\sum_{i}{\delta \left({\mathit{OKS}_{i}}>\tau \right)}}{\sum_{i}{\mathit{1} }},
\end{equation}
where we use standard $\mathit{AP}_{\tau}$, which measures prediction precision of a model given a specific threshold ${\tau}$. In order to provide a comprehensive and nuanced evaluation of the model's performance, we report average precision across various thresholds. Specifically, we report $\mathit{AP}^{50}$ ($\mathit{AP}$ at $\tau=0.5$), $\mathit{AP}^{75}$ ($\mathit{AP}$ at $\tau=0.75$), $\mathit{AP}$ (the mean $\mathit{AP}$ across 10 $\tau$ positions, where $\tau=\{0.5,0.55,...,0.95\}$), $\mathit{AP}^{S}$ (for small vessels with edge radius within the range of $(0,2]$), $\mathit{AP}^{M}$ (for medium vessels with edge radius within the range of $(2,3]$), $\mathit{AP}^{L}$ (for large vessels with edge radius greater than 3), $\mathit{AP}^{k1}$ ($\mathit{AP}$ for $\mathit{KP_{1}}$), $\mathit{AP}^{k2}$ ($\mathit{AP}$ for $\mathit{KP_{2}}$), $\mathit{E_d}$ (mean $\mathit{E}_{d_{k}}$), $\mathit{E_d}^{k1}$, $\mathit{E_d}^{k2}$ ($\mathit{E_d}$ for $\mathit{KP_{1}}$, $\mathit{KP_{2}}$).

\subsection{Results} 

% !TEX root = ../top.tex
% !TEX spellcheck = en-US

\begin{table}[tb]
	\caption{\textbf{Keypoint detection performance on the PTR dataset.} \textbf{$\mathit{UNet^{1}}$}: The input is one subvolume with both $\mathit{KP_{1}}$ and $\mathit{KP_{2}}$ components; \textbf{$\mathit{UNet^{2}}$}: The inputs are two concatenated subvolumes of $\mathit{KP_{1}}$ and $\mathit{KP_{2}}$ components; Metrics are expressed in percentage (\%) format.}
	\label{tab:keypoint_point_detection}
	\centering
 % p{1.5cm}<{\centering}
    	\begin{tabular*}{\linewidth}{c| p{1.3cm}<{\centering} |@{\extracolsep{\fill}}c|cc|cc|ccc|c|cc}
    		\toprule
    		Task & Method & $\mathit{AP}$ & $\mathit{AP}^{k1}$ & $\mathit{AP}^{k2}$ & $\mathit{AP}^{50}$ & $\mathit{AP}^{75}$ & $\mathit{AP}^{S}$ & $\mathit{AP}^{M}$ & $\mathit{AP}^{L}$ & $\mathit{E_d}$ & $\mathit{E_d}^{k1}$ & $\mathit{E_d}^{k2}$\\
    		\midrule
            \multirow{2}{*}{Airway} & $\mathit{UNet^{1}}$ & 80.89 & 79.23 & 86.32 & 94.21 & 90.47 & 75.15 & 85.32 & 80.02 & 18.81 & 15.39 & 22.07\\
            & $\mathit{UNet^{2}}$ & \bf 87.18 & 83.80 & 90.88 & 98.48 & 94.89 & 79.56 & 90.29 & 93.17 & \bf 28.54 & 23.47 & 33.37\\
            
            \midrule
            
            \multirow{2}{*}{Artery} & $\mathit{UNet^{1}}$ & 71.32 & 70.19 & 81.90 & 85.99 & 78.89 & 62.98 & 81.11 & 68.97 & 16.71 & 12.55 & 20.65\\
            & $\mathit{UNet^{2}}$ & \bf 80.58 & 77.46 & 87.07 & 94.09 & 86.85 & 70.45 & 88.31 & 84.25 & \bf 25.49 & 20.83 & 29.90\\
             
            \midrule       

            \multirow{2}{*}{Vein} & $\mathit{UNet^{1}}$ & 69.10 & 69.09 & 79.79 & 82.80 & 76.97 & 59.17 & 78.25 & 69.38 & 15.49 & 12.26 & 18.54\\
            & $\mathit{UNet^{2}}$ & \bf 78.78 & 76.27 & 85.45 & 93.23 & 84.95 & 67.26 & 85.71 & 85.22 & \bf 24.40 & 20.81 & 27.79\\
            % & $\mathit{UNet^{2}}$ & \bf 84.61 & 79.66 & 91.95 & 95.23 & 90.38 & 75.19 & 91.23 & 88.92 & \bf 33.75 & 21.77 & 45.09\\
    		
    		\bottomrule
    	\end{tabular*}
\end{table}

% \begin{wraptable}{r}{5.5cm}
% 	\caption{\textbf{Repair results on Distorted Golden Standards.} We compare three four, \ie standard segmentation (Seg-FCN and Seg-UNet), NeAR w/ shape only (S) and NeAR w/ shape and appearence (S+A), and report Dice Similarity Coefficient (DSC)(\%) and Normalized Surface Dice (NSD)(\%) for each method.}
% 	\label{tab:distorted_golden_standards}
% 	\centering
%     	\begin{tabular}{l|c|c}
%     		\toprule
%     		Methods & DSC & NSD \\
%     		\midrule
%     		Seg-FCN & 81.11 & 89.26  \\
%     		Seg-UNet & 81.00 & 89.13  \\
%     		NeAR (S) & 79.38 & 87.88  \\
%     		NeAR (S+A) & \bf 81.33 & \bf 91.21  \\
%     		\bottomrule
%     	\end{tabular}
% \end{wraptable}

To analyze the performance of our methods on topology repairing of disconnected pulmonary airways and vessels, we report several methods on the proposed PTR dataset, as shown in Table~\ref{tab:keypoint_point_detection}. 
The keypoint heatmap visualization is provided in the supplementary materials.

The study demonstrates that the two-channel 3D-UNet model surpasses the performance of the one-channel counterpart on airway and vessel segmentation tasks. Specifically, the two-channel model yields significant improvements in $\mathit{AP}$ of approximately 7\%, 9\%, and 15\% for airway, artery, and vein tasks, respectively. Additionally, the two-channel model achieves the highest performance on all evaluation metrics for all three tasks. These results suggest that the separation of $\mathit{KP_{1}}$ and $\mathit{KP_{2}}$ components as two-channel input can effectively improve their interaction in multiple feature levels, leading to improved performances. This is likely due to the high correlation between these two keypoints throughout the topological structure. However, detecting $\mathit{KP_{1}}$ was significantly challenging due to the random selection of cropping center points during data sampling, leading to a weaker performance for $\mathit{E_d}$ and $\mathit{AP}$ metrics. Additionally, the sparse distribution of keypoints on small pulmonary vessels posed a considerable challenge for capturing subtle features. Notably, the two-channel networks exhibited superior performance over one-channel methods by a substantial margin, which emphasizes the advantages of separating the two components. In the future work, it will be beneficial to design models that capture this characteristic. 
% !TEX root = ../top.tex
% !TEX spellcheck = en-US

\section{Conclusion}

In this study, we introduce a data-driven post-processing approach that addresses the challenge of disconnected pulmonary tubular structures, which is crucial for the diagnosis and treatment of pulmonary diseases. The proposed approach utilizes the newly created Pulmonary Tree Repairing (PTR) dataset, comprising 800 complete 3D models of pulmonary structures and synthetic disconnected data. A two-channel simple yet effective neural network is trained to detect keypoints that bridge disconnected components, utilizing a training data synthesis pipeline that generates disconnected data from complete pulmonary structures. Our approach yields promising results and holds great potential for clinical applications. {While our study primarily focuses on addressing the disconnection issue, we recognize that more complex scenarios, such as handling multiple disconnected components, distinguishing between arteries and veins, and implementing our method in real-world settings, require further investigation in future work. Point or implicit representations~\cite{yang2021ribseg,yang2022implicitatlas,yang2022neural} learning the geometric structures have high potentials in this application.} 

\subsubsection{Acknowledgment.}
This research was supported by Australian Government Research Training Program (RTP) scholarship, and supported in part by a Swiss National Science Foundation grant.

%%%%%%%%% REFERENCES

%%%%%%%%% Supplementary
\title{\textit{Supplementary Materials}\\
Topology Repairing of Disconnected Pulmonary Airways and Vessels: Baselines and a Dataset}

\titlerunning{Topology Repairing of Pulmonary Airways and Vessels}

% \institute{}
\author{Ziqiao Weng\inst{1} \and
Jiancheng Yang\inst{2}\thanks{Corresponding author: Jiancheng Yang (jiancheng.yang@epfl.ch).} \and
Dongnan Liu\inst{1} \and
Weidong Cai\inst{1}}
\authorrunning{Z. Weng et al.}
\institute{School of Computer Science, University of Sydney, Sydney, Australia \and
Computer Vision Laboratory, Swiss Federal Institute of Technology Lausanne (EPFL), Lausanne, Switzerland
}

\maketitle              % typeset the header of the contribution

\appendix

\setcounter{figure}{0}
\renewcommand{\thefigure}{A\arabic{figure}}

\begin{algorithm}[H]
	\caption{Disconnection Keypoint Detection for Whole Volume}
	\label{algo:inference}
	% \small
	\SetAlgoLined
	\KwInput{ $\boldsymbol{X} \in \mathbb{R}^{D\times H\times W}$ \tcp{3D binarized input volume} \\
    \qquad
    $\boldsymbol{\mathit{f} \left ( ,\omega ^{\ast }  \right ) }$ \tcp{model with learned parameters $\omega ^{\ast }$} \\
    \qquad
    $\boldsymbol{\mathit{T}}$ \tcp{number of crops for each component} \\
    \qquad
    $\boldsymbol{\mathit{K}}$ \tcp{number of keypoint}
    }

	\KwOutput{ $\boldsymbol{P} \in \mathbb{R}^{k\times 3}$, $\mathit{P}= \left \{ P_{1},..., P_{K} \right \}$ \tcp{predicted keypoint coordinates}
    }
    
    \KwInitialize{
    $zeros(\boldsymbol{H}) \in \mathbb{R}^{K\times D\times H\times W}$, $\mathit{H}= \left \{ H_{1},..., H_{K} \right \}$ \tcp{predicted keypoint heatmaps}
    }
    
    % \nl $C=\mathit{ConnectedComponent}(\boldsymbol{X}, filter\_noises=True, filter\_background=True)$
    \nl $C=\mathit{ConnectedComponent}(\boldsymbol{X}, \mathit{filter\_noises}=True)$ 
    
    \nl \tcp{candidates of disconnected component} $C=\mathit{RemoveLargestComponent}(C)$, $\mathit{C}= \left \{ C_{1},..., C_{M} \right \}$ 
    
    \nl \For{$C_{i}=C_{1},..., C_{M}$}    
        { 
        	\For{$t=1,..., T$}    
            { 
        	$P_{t}^{'}=random\_sample(C_{i}) \in \mathbb{R}^{3}$\\
         
            $X_{t}=crop(\boldsymbol{X}, center=P_{t}^{'}) \in \mathbb{R}^{D^{'}\times H^{'}\times W^{'}}$ \\
            
            $\hat{H_{t}}=\mathit{f}(X_{t},\omega ^{\ast }) \in \mathbb{R}^{3\times D^{'}\times H^{'}\times W^{'}}$ \\
            \tcp{fill $H$ with $\hat{H_{t}}$ according to $P_{t}^{'}$}
            $\boldsymbol{H}[get\_location(P_{t}^{'})] \mathrel{+}=\hat{H_{t}}$ 
            
            }
        }

    \nl \For{$k=1,..., K$} 
    {
        $P_{k}=argmax(H_{k}) \in \mathbb{R}^{3}$\\
    
    }
        
	\nl Return $\boldsymbol{P} \in \mathbb{R}^{k\times 3}, \mathit{P}= \left \{ P_{1},..., P_{K} \right \}$
\end{algorithm}

\begin{figure}[htb]
	\centerline{\includegraphics[width=\linewidth]{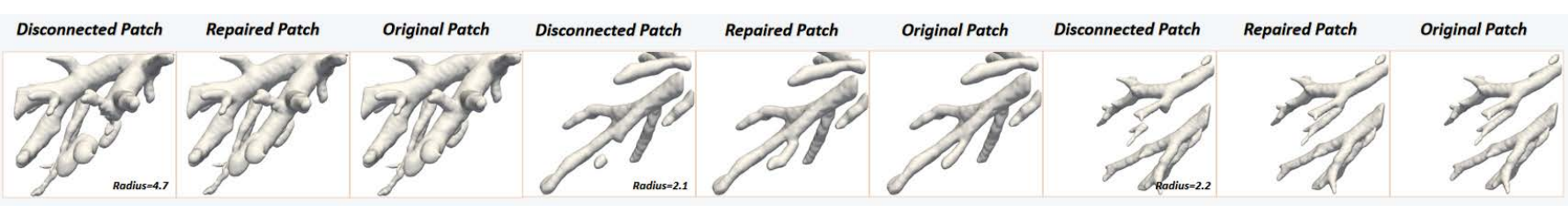}}
    \captionsetup{font={scriptsize}} 
	\caption{\textbf{Repairing of Disconnected Airways or Vessels by Linking Predicted Keypoints.} Here we show a simple and rudimentary method of repairing disconnections by linking the two key points with a cylinder of radius $\mathit{R}$ and place a hemisphere of radius $\mathit{R}$ at key point. Notably, the value of $\mathit{R}$ is set equal to the radius of the corresponding vessel or airway.}
	\label{fig:connect}
\end{figure}
% !TEX root = ../top.tex
% !TEX spellcheck = en-US
% [!htb]

\begin{figure}[htb]
\centering
% \subfigure{\scriptsize{(a) Airway}}{
\subfigure{}{
    \begin{minipage}[b]{0.8\linewidth} %0.23为minipage的宽度，可以调节子图间的距离
    \centering
    \includegraphics[width=\linewidth]{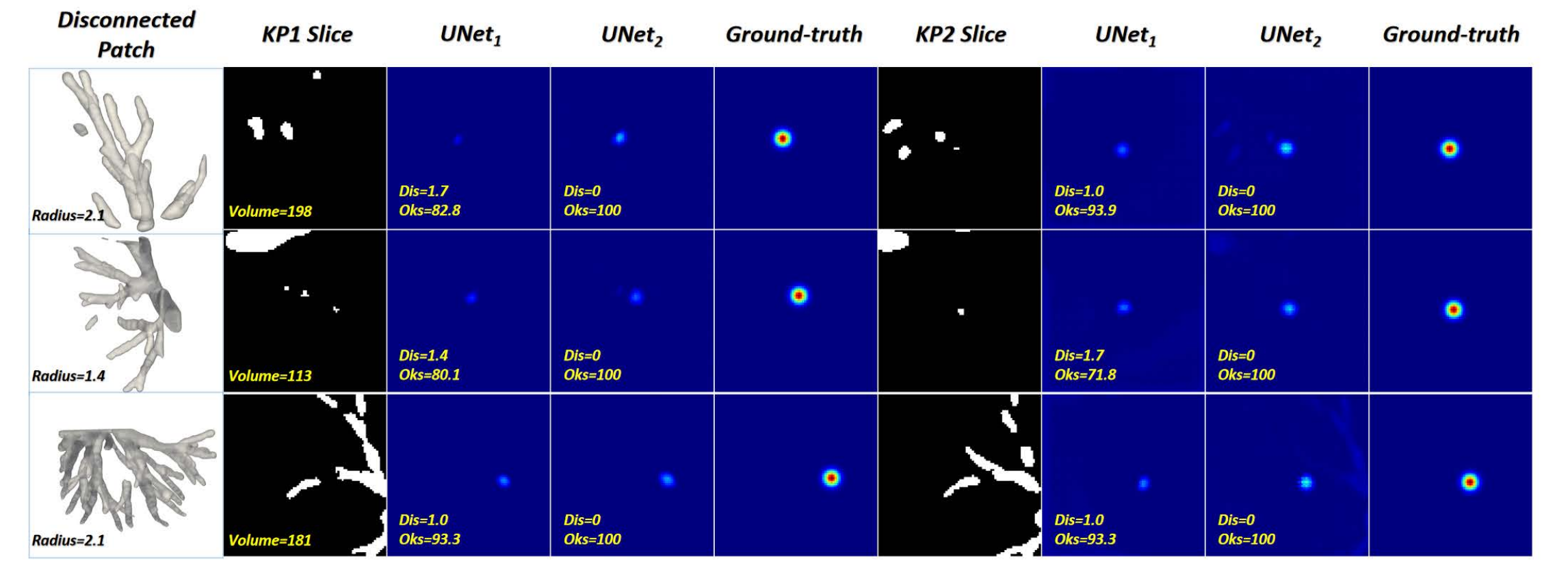}\vspace{0.1pt} %图片的宽度、路径和垂直间距
    %\vspace要紧跟在对应的includegraphics，不然得不到想要的结果
    \end{minipage}
}
% \quad %退一格
% \qquad %退两格,调节子图间的距离
% \subfigure{\scriptsize{(b) Artery}}{
\subfigure{}{
    \begin{minipage}[b]{0.8\linewidth}
    \centering
    \includegraphics[width=\linewidth]{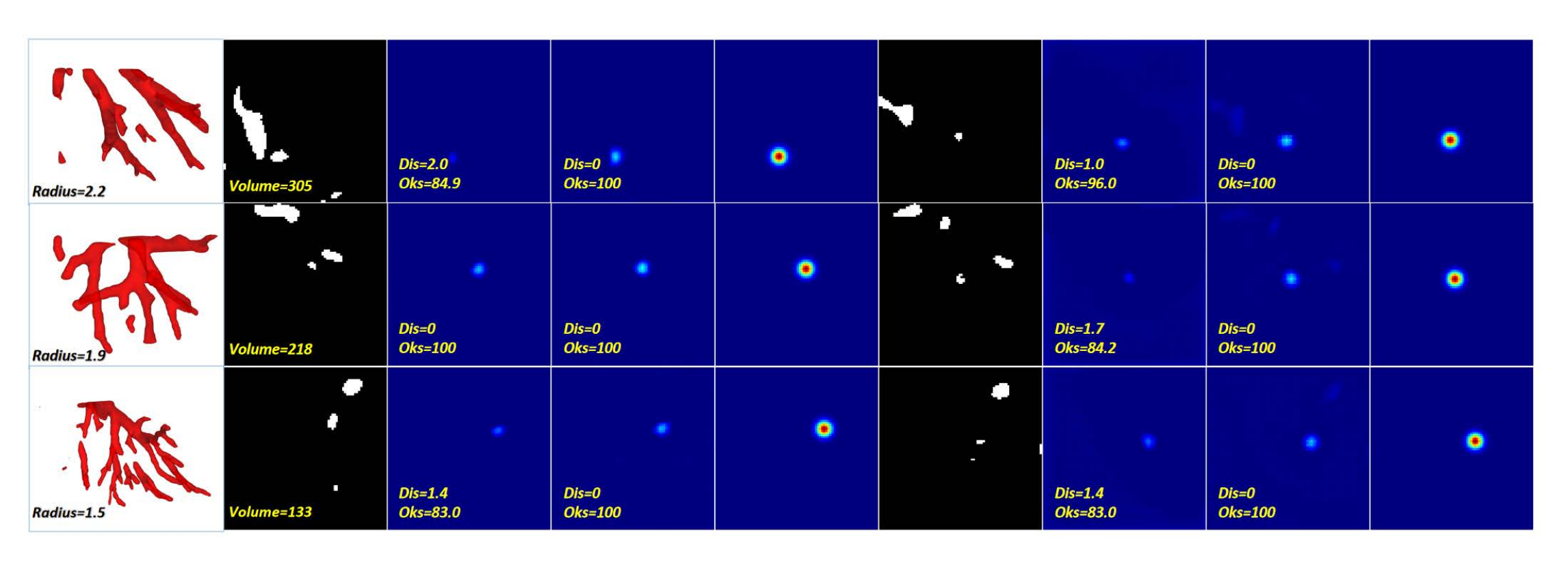}\vspace{0.1pt}
    \end{minipage}
}

% \subfigure{\scriptsize{(c) Vein}}{
\subfigure{}{
    \begin{minipage}[b]{0.8\linewidth}
    \centering
    \includegraphics[width=\linewidth]{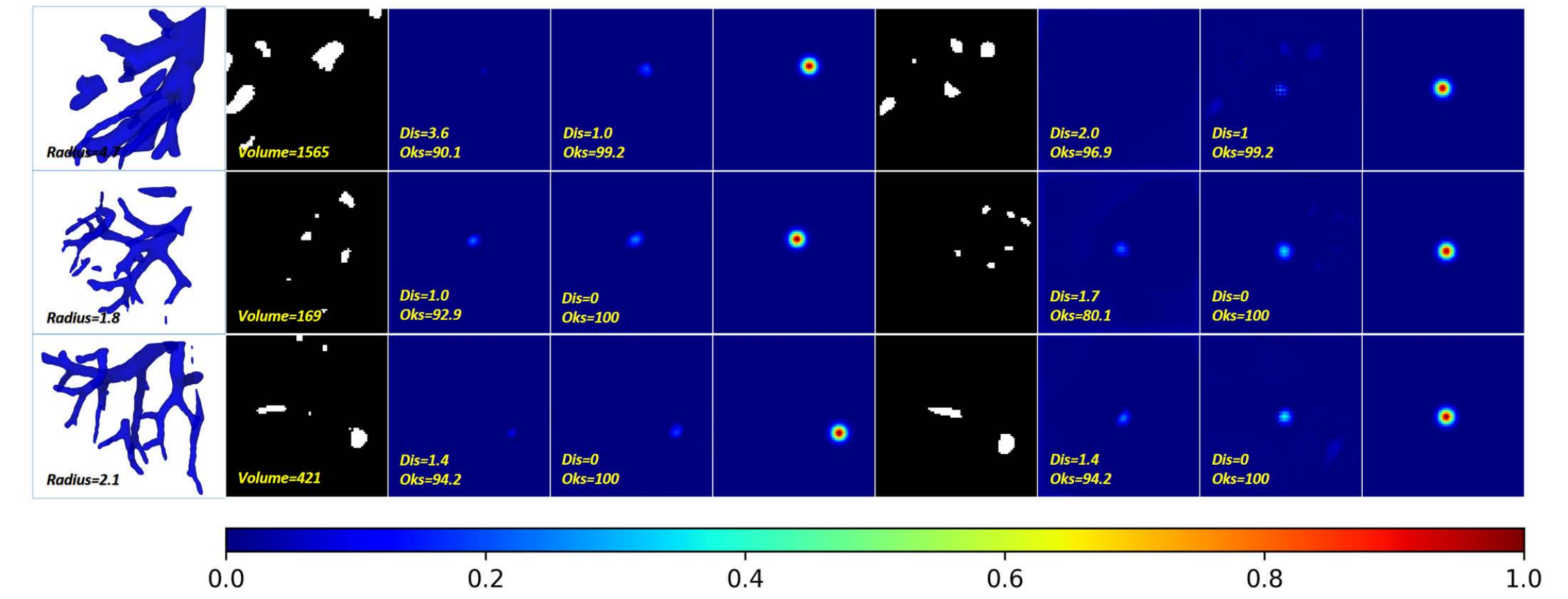}
    \end{minipage}
}
% \captionsetup{font={scriptsize}} 
\caption{\textbf{Heatmap Visualizations. Top: Airway. Middle: Artery. Bottom: Vein.} We show a visual comparison of the performance of two UNet models in predicting keypoints for airway, artery and vein. The first column denotes the 3D view of disconnected subvolume. The second column displays a 2D slice (x-y plane) of the subvolume corresponding to $\mathit{KP_{1}}$. The next two columns illustrate the predicted heatmap slice of $\mathit{KP_{1}}$ by the two UNet models, followed by the ground-truth heatmap slice of $\mathit{KP_{1}}$ in the fourth column. The final four columns follow the same visualization as $\mathit{KP_{1}}$ for $\mathit{KP_{2}}$. The vessel or airway's radius and volume size are reported as \textbf{Radius} and \textbf{Volume}. \textbf{Oks} (in (\%) format) and \textbf{Dis} represent the OKS and Euclidean distance between the predicted and ground-truth keypoints, respectively.}
\vspace{-20pt} 
\end{figure}

% \begin{figure}[!htb]
% 	\includegraphics[width=\linewidth]{fig/SM-heatmap-airway.png}
% 	\caption{\textbf{Heatmap Visualization of Airway Task.}}
% 	\label{fig:SM-heatmap-airway}
% \end{figure}

% \begin{figure}[!htb]
% 	\includegraphics[width=\linewidth]{fig/SM-heatmap-artery.png}
% 	\caption{\textbf{Heatmap Visualization of Artery Task.} }
% 	\label{fig:SM-heatmap-artery}
% \end{figure}


\begin{thebibliography}{10}
\providecommand{\url}[1]{\texttt{#1}}
\providecommand{\urlprefix}{URL }
\providecommand{\doi}[1]{https://doi.org/#1}

\bibitem{SimpleITK1}
Beare, R., Lowekamp, B., Yaniv, Z.: Image segmentation, registration and
  characterization in r with simpleitk. Journal of Statistical Software
  \textbf{86} (08 2018). \doi{10.18637/jss.v086.i08}

\bibitem{bumgarner2022vesselvio}
Bumgarner, J.R., Nelson, R.J.: Open-source analysis and visualization of
  segmented vasculature datasets with vesselvio. Cell Reports Methods
  \textbf{2}(4),  100189 (2022).
  \doi{https://doi.org/10.1016/j.crmeth.2022.100189},
  \url{https://www.sciencedirect.com/science/article/pii/S2667237522000443}

\bibitem{fetita2004pulmonary}
Fetita, C.I., Pr{\^e}teux, F., Beigelman-Aubry, C., Grenier, P.: Pulmonary
  airways: 3-d reconstruction from multislice ct and clinical investigation.
  IEEE Transactions on Medical Imaging  \textbf{23}(11),  1353--1364 (2004)

\bibitem{garcia2021automatic}
Garcia-Uceda, A., Selvan, R., Saghir, Z., Tiddens, H.A., de~Bruijne, M.:
  Automatic airway segmentation from computed tomography using robust and
  efficient 3-d convolutional neural networks. Scientific Reports
  \textbf{11}(1),  1--15 (2021)

\bibitem{isensee2021nnu}
Isensee, F., Jaeger, P.F., Kohl, S.A., Petersen, J., Maier-Hein, K.H.: nnu-net:
  a self-configuring method for deep learning-based biomedical image
  segmentation. Nature methods  \textbf{18}(2),  203--211 (2021)

\bibitem{Kuang2022WhatMF}
Kuang, K., Zhang, L., Li, J., Li, H., Chen, J., Du, B., Yang, J.: What makes
  for automatic reconstruction of pulmonary segments. In: Conference on Medical
  Image Computing and Computer Assisted Intervention. pp. 495--505. Springer
  (2022)

\bibitem{LuoSWAHR}
Luo, Z., Wang, Z., Huang, Y., Wang, L., Tan, T., Zhou, E.: Rethinking the
  heatmap regression for bottom-up human pose estimation. In: Conference on
  Computer Vision and Pattern Recognition (2021)

\bibitem{DeepVessel}
Pan, C., Qi, B., Zhao, G., Liu, J., Fang, C., Zhang, D., Li, J.: Deep 3d vessel
  segmentation based on cross transformer network. 2022 IEEE International
  Conference on Bioinformatics and Biomedicine pp. 1115--1120 (2022)

\bibitem{park2021deep}
Park, J., Hwang, J., Ryu, J., Nam, I., Kim, S.A., Cho, B.H., Shin, S.H., Lee,
  J.Y.: Deep learning based airway segmentation using key point prediction.
  Applied Sciences  \textbf{11}(8), ~3501 (2021)

\bibitem{qin2021learning}
Qin, Y., Zheng, H., Gu, Y., Huang, X., Yang, J., Wang, L., Yao, F., Zhu, Y.M.,
  Yang, G.Z.: Learning tubule-sensitive cnns for pulmonary airway and
  artery-vein segmentation in ct. IEEE Transactions on Medical Imaging
  \textbf{40}(6),  1603--1617 (2021)

\bibitem{rahaghi2016pulmonary}
Rahaghi, F., Ross, J., Agarwal, M., Gonz{\'a}lez, G., Come, C., Diaz, A.,
  Vegas-S{\'a}nchez-Ferrero, G., Hunsaker, A., Est{\'e}par, R.S.J., Waxman, A.,
  et~al.: Pulmonary vascular morphology as an imaging biomarker in chronic
  thromboembolic pulmonary hypertension. Pulmonary circulation  \textbf{6}(1),
  70--81 (2016)

\bibitem{saji2022segmentectomy}
Saji, H., Okada, M., Tsuboi, M., Nakajima, R., Suzuki, K., Aokage, K., Aoki,
  T., Okami, J., Yoshino, I., Ito, H., et~al.: Segmentectomy versus lobectomy
  in small-sized peripheral non-small-cell lung cancer (jcog0802/wjog4607l): a
  multicentre, open-label, phase 3, randomised, controlled, non-inferiority
  trial. The Lancet  \textbf{399}(10335),  1607--1617 (2022)

\bibitem{sun2019deep}
Sun, K., Xiao, B., Liu, D., Wang, J.: Deep high-resolution representation
  learning for human pose estimation. In: Conference on Computer Vision and
  Pattern Recognition (2019)

\bibitem{tetteh2020deepvesselnet}
Tetteh, G., Efremov, V., Forkert, N.D., Schneider, M., Kirschke, J., Weber, B.,
  Zimmer, C., Piraud, M., Menze, B.H.: Deepvesselnet: Vessel segmentation,
  centerline prediction, and bifurcation detection in 3-d angiographic volumes.
  Frontiers in Neuroscience p.~1285 (2020)

\bibitem{andongW}
Wang, A., Tam, T., Poon, H., Yu, K.C., Lee, W.N.: Naviairway: a
  bronchiole-sensitive deep learning-based airway segmentation pipeline for
  planning of navigation bronchoscopy. arXiv Preprint  (03 2022).
  \doi{10.36227/techrxiv.19228296}

\bibitem{WangJD}
Wang, J., Sun, K., Cheng, T., Jiang, B., Deng, C., Zhao, Y., Liu, D., Mu, Y.,
  Tan, M., Wang, X., Liu, W., Xiao, B.: Deep high-resolution representation
  learning for visual recognition. IEEE Transactions on Pattern Analysis and
  Machine Intelligence  (2019)

\bibitem{Wittenberg}
Wittenberg, R., Berger, F., Peters, J., Weber, M., Hoorn, F., Beenen, L., van
  Doorn, M., Schuppen, J., Zijlstra, I., Prokop, M., Schaefer-Prokop, C.: Acute
  pulmonary embolism: Effect of a computer-assisted detection prototype on
  diagnosis-an observer study. Radiology  \textbf{262},  305--13 (01 2012).
  \doi{10.1148/radiol.11110372}

\bibitem{xiao2018simple}
Xiao, B., Wu, H., Wei, Y.: Simple baselines for human pose estimation and
  tracking. In: European Conference on Computer Vision (2018)

\bibitem{yang2021ribseg}
Yang, J., Gu, S., Wei, D., Pfister, H., Ni, B.: Ribseg dataset and strong point
  cloud baselines for rib segmentation from ct scans. In: Conference on Medical
  Image Computing and Computer Assisted Intervention. pp. 611--621. Springer
  (2021)

\bibitem{yang2022neural}
Yang, J., Shi, R., Wickramasinghe, U., Zhu, Q., Ni, B., Fua, P.: Neural
  annotation refinement: Development of a new 3d dataset for adrenal gland
  analysis. In: Conference on Medical Image Computing and Computer Assisted
  Intervention. pp. 503--513. Springer (2022)

\bibitem{yang2022implicitatlas}
Yang, J., Wickramasinghe, U., Ni, B., Fua, P.: Implicitatlas: learning
  deformable shape templates in medical imaging. In: Conference on Computer
  Vision and Pattern Recognition. pp. 15861--15871 (2022)

\bibitem{yaniv2018simpleitk}
Yaniv, Z., Lowekamp, B.C., Johnson, H.J., Beare, R.: Simpleitk image-analysis
  notebooks: a collaborative environment for education and reproducible
  research. Journal of digital imaging  \textbf{31}(3),  290--303 (2018)

\bibitem{yao2021one}
Yao, Q., Quan, Q., Xiao, L., Kevin~Zhou, S.: One-shot medical landmark
  detection. In: Conference on Medical Image Computing and Computer Assisted
  Intervention. pp. 177--188. Springer (2021)

\bibitem{yu2021heatmap}
Yu, B., Tao, D.: Heatmap regression via randomized rounding. IEEE Transactions
  on Pattern Analysis and Machine Intelligence  (2021)

\bibitem{zhang2023multi}
Zhang, M., Wu, Y., Zhang, H., Qin, Y., Zheng, H., Tang, W., Arnold, C., Pei,
  C., Yu, P., Nan, Y., et~al.: Multi-site, multi-domain airway tree modeling
  (atm'22): A public benchmark for pulmonary airway segmentation. arXiv
  Preprint  (2023)

\bibitem{zhao20183d}
Zhao, W., Yang, J., Sun, Y., Li, C., Wu, W., Jin, L., Yang, Z., Ni, B., Gao,
  P., Wang, P., et~al.: 3d deep learning from ct scans predicts tumor
  invasiveness of subcentimeter pulmonary adenocarcinomas. Cancer research
  \textbf{78}(24),  6881--6889 (2018)

\bibitem{zhao2023invasiveness}
Zhao, Z.R., Yu, Y.H., Lin, Z.C., Ma, D.H., Lin, Y.B., Hu, J., Luo, Q.Q., Li,
  G.F., Chen, C., Yang, Y.L., et~al.: Invasiveness assessment by artificial
  intelligence against intraoperative frozen section for pulmonary nodules
  $\leq$ 3 cm. Journal of Cancer Research and Clinical Oncology pp.~1--7 (2023)

\end{thebibliography}
\end{document}